\documentclass[11pt,twoside]{article}
\usepackage{./asp2014}

\aspSuppressVolSlug
\resetcounters

\begin{document}

\title{Quantifying the gas inside dust cavities in transitional disks: implications for young planets}
\author{E.F. van Dishoeck,$^{1,2}$ N. van der Marel,$^1$ S. Bruderer,$^2$, and P. Pinilla$^1$}
\affil{$^1$Leiden Observatory, Leiden University, the Netherlands; \email{ewine@strw.leidenuniv.nl}}
\affil{$^2$Max Planck Institute for Extraterrestrial Physics, Garching, Germany}

\paperauthor{E.F. van Dishoeck}{ewine@strw.leidenuni.nl}{ORCID_Or_Blank}{Leiden Observatory}{Leiden University}{Leiden}{ZH}{2300 RA}{The Netherlands}
\paperauthor{N. van der Marel}{nmarel@strw.leidenuni.nl}{ORCID_Or_Blank}{Leiden Observatory}{Leiden University}{Leiden}{ZH}{2300 RA}{The Netherlands}
\paperauthor{S. Bruderer}{simonbruderer@gmail.com}{ORCID_Or_Blank}{MPE}{}{Garching}{}{85748}{Germany}
\paperauthor{P. Pinilla}{pinilla@strw.leidenuni.nl}{ORCID_Or_Blank}{Leiden Observatory}{Leiden University}{Leiden}{ZH}{2300 RA}{The Netherlands}

\begin{abstract}
  ALMA observations of a small sample of transitional disks with large
  dust cavities observed in Cycle 0 and 1 are summarized. The gas and
  dust surface density structures are inferred from the continuum and
  $^{12}$CO, $^{13}$CO and C$^{18}$O line data using the DALI
  physical-chemical code. Thanks to its ability to self-shield, CO can
  survive inside dust cavities in spite of being exposed to intense UV
  radiation and can thus be used as a probe of the gas
  structure. Modeling of the existing data shows that gas is present
  inside the dust cavities in all cases, but at a reduced level
  compared with the gas surface density profile of the outer disk. The
  gas density decrease inside the dust cavity radius by factors of up
  to $10^4$ suggests clearing by one or more planetary-mass
  companions. The accompanying pressure bumps naturally lead to
  trapping of the mm-sized dust grains observed in the ALMA images.
\end{abstract}

\section{Introduction}

Transitional disks with large inner dust cavities are excellent
laboratories for studying disk evolution during the planet-forming
stage (Brown et al.\ 2007, Andrews et al.\ 2011, Espaillat et al.\
2014). Little is known about the cold molecular gas inside dust
cavities, yet this gas significantly affects planet formation through
gas-grain dynamics and planetary migration whereas its chemistry
controls the composition of gas-giant atmospheres.  Moreover, the gas
surface density structure puts constraints on the origin of the
cavity, in particular whether grain growth or photoevaporation can
explain the data or whether a companion is needed. With its tremendous
sensitivity, ALMA has opened up the possibility to directly detect and
characterize the colder molecular gas inside dust cavities.

In this paper, we briefly summarize the results on the gas and dust
structures from our ALMA Cycle 0 and 1 observations of 4 transitional
disks: Oph IRS 48 observed in Band 9 (van der Marel et al.\ 2013,
hereafter vdM13), and SR21, HD 135344B and DoAr 44 taken in Band 7
(van der Marel et al., in prep. (vdM15b)). The spatial resolution is
typically 0.20--0.25$''$, corresponding to 15 AU in radius at the
distance of Ophiuchus (125 pc). We also use archival Band 9 data for
the latter three sources from P\'erez et al. (2014, P14), summarized
in van der Marel et al.\ (2015; vdM15a). Data on $^{12}$CO, $^{13}$CO
and C$^{18}$O $J$=6--5 and/or 3--2 isotopologs are available.  These
sources have been studied with CO ro-vibrational lines at infrared
wavelengths as well, providing independent information of gas inside
the dust gap (Pontoppidan et al.\ 2008).

\articlefigure[width=.8\textwidth]{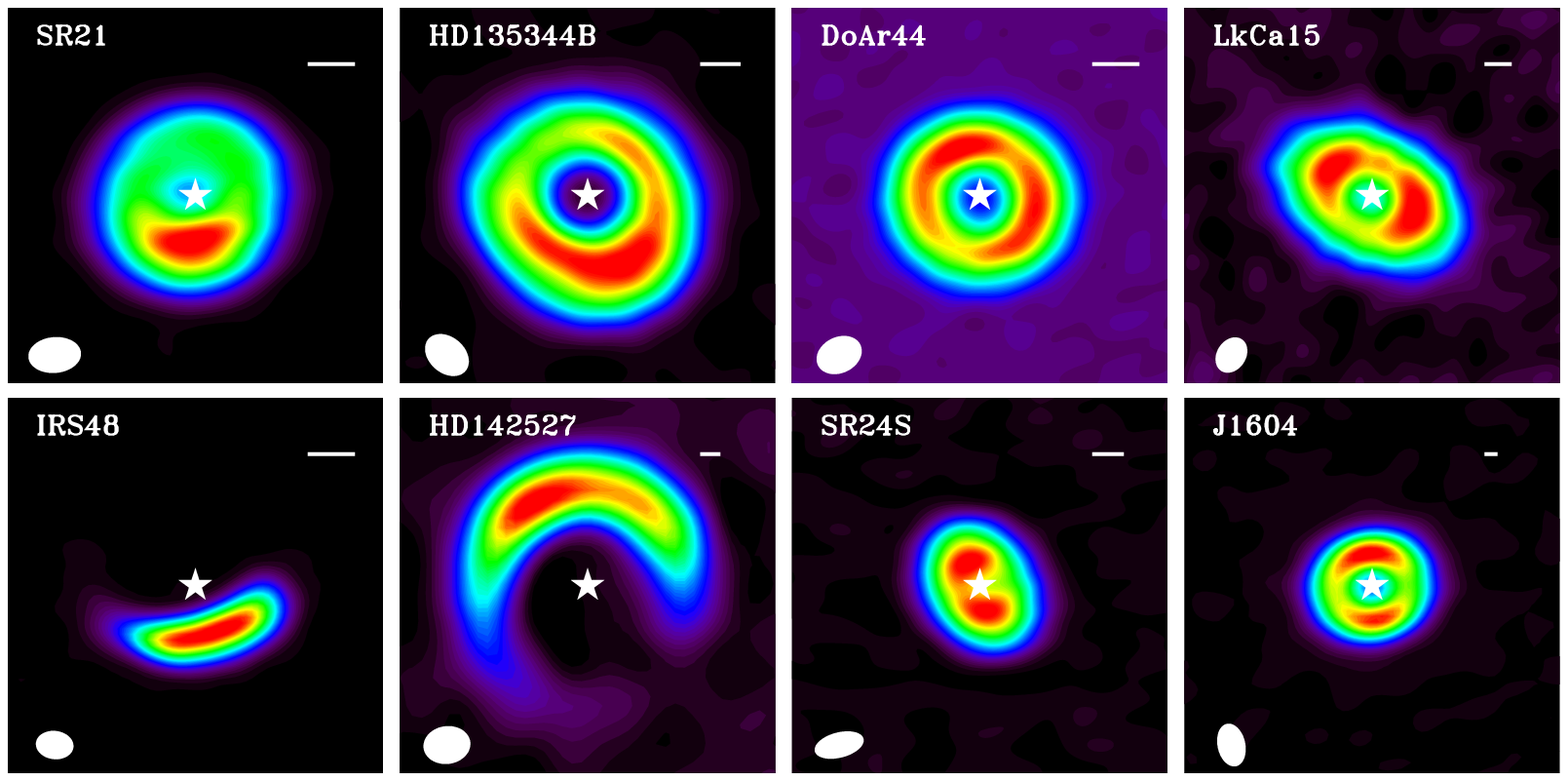}{fig1}{ALMA
  continuum images of 8 transitional disks; the colors are scaled to
  the peak emission in each image but the azimuthal contrast varies greatly from
  disk to disk (see text).  {\bf Top:} SR 21 (B9; P14), HD 135344B (B9;
  P14), DoAr44 (B7; vdM15b), LkCa15 (B9; vdM15a); {\bf Bottom:}
  IRS 48 (B9; vdM15b), HD 142527 (B7; Cassasus et al.\ 2013), SR 24S
  (B9; vdM15a), J1604-2130 (B7, Zhang et al.\ 2014).  The horizontal
  white bar indicates a 30 AU scale.}

\section{Dust: images and models}

Figure 1 presents a gallery of the millimeter-sized dust emission from
a number of transitional disks to put our four sources in context. It
is clear that there is a wide variety of dust structures: some souces
such as DoAr44 and J1604-2130 show very symmetric (inclined) dust
rings with only a $\sim$20\% variation in brightness across the
ring. Other sources such as HD 135344B and SR 21 have lopsided
structures with a factor of $\sim$2 variation. The most extreme cases
are Oph IRS 48 and HD 142527 which show highly asymmetric dust
continuum emission with azimuthal contrasts of $>$100 and $\sim$30,
respectively.  In contrast, the small micron-sized dust grains are
distributed much more symmetrically along the IRS 48 ring (Geers et
al.\ 2007, vdM13).

In general, these dust structures are well modeled by dust traps
caused by a gas pressure bump in which the dust particles accumulate
and grow (e.g., Barge \& Sommeria 1995, Klahr \& Henning 1997, Pinilla
et al.\ 2012). One exciting possibility is that the pressure bumps are
the result of planet-disk interactions and thus allow characterization
of the unseen planets: the efficiency of the dust trapping (largest
grain size produced) and the location of the dust depend on the orbit
and mass of the companion, as well as on the turbulence in the disk.
However, alternative models such as pressure bumps triggered by
instabilities at the edges of dead zones have also been invoked and
dust images alone cannot readily distinguish between these options
(e.g., Reg\'aly et al.\ 2013, Flock et al.\ 2015).

The strong azimuthal asymmetries for IRS 48 and HD 142527 can be
explained by the presence of a long-lived ($\sim 10^5$ yr) vortex
caused by a Rossby wave shearing instability at the steep gas
density edges (e.g., Ataiee et al.\ 2013). Such vortices develop only
under certain conditions and require low disk viscosity. Which
particles get trapped in the vortices depends on their Stokes number,
which in turn is tied to the local gas density.

\articlefigure[width=.9\textwidth]{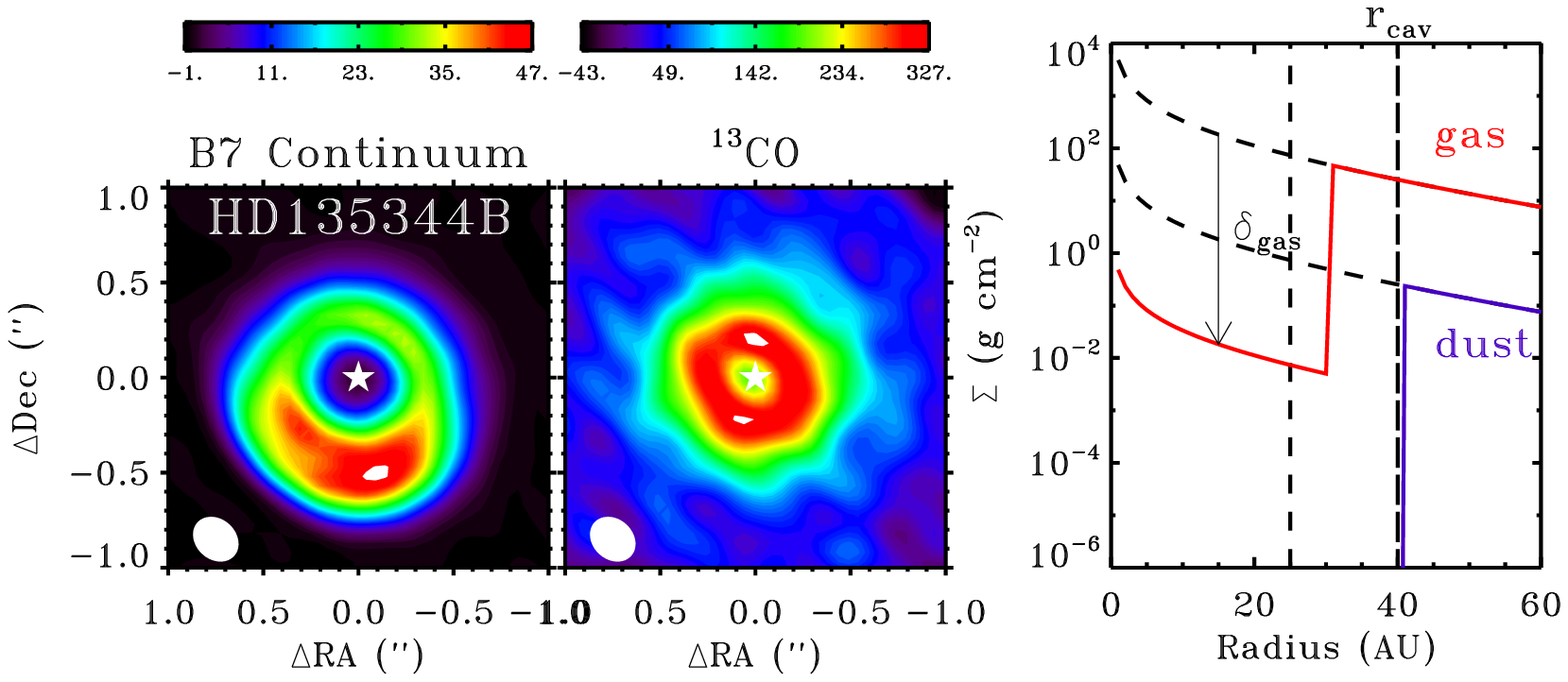}{fig2}{Comparison
  of Band 7 dust (left) and gas ($^{13}$CO $J$=3-2, middle) images for
  HD135344B, together with the inferred dust and gas surface density
  profiles (vdM15a,vdM15b).}

\section{Gas: images and models}

In contrast with the dust, the gas distribution is generally much more
symmetric along the ring, even for sources with extreme dust traps
(Fig.\ 2). Moreover, gas is clearly detected inside the hole,
demonstrating that gas and dust have very different distributions in
these disks (e.g., Bruderer et al.\ 2014, P14, Zhang et al.\ 2014,
Perez et al.\ 2015, vdM15a).

To quantify the gas surface density, the thermo-chemical code DALI
developed by Bruderer et al.\ (2012, 2013) is used.  This code
computes the dust temperature at each position in the disk through
full continuum radiative transfer from UV to centimeter wavelengths,
as well as the heating-cooling balance of the gas together with the
chemistry. The non-LTE molecular excitation and line radiative
transfer is then calculated using the gas temperature and molecular
abundances as inputs and finally ray-traced to compute line images
for the disk's inclination.

The baseline model assumes a parametric gas density model cf.\ Andrews
et al.\ (2011), which has a small inner dust disk and a dust cavity
out to $r_{\rm cav}$ but no break in the gas distribution. Grain
growth and settling are taken into account through a simple
parametrization of the dust properties. Subsequently the amount of gas
inside the dust cavity is lowered by factors of 10$^x$, with $x=1-6$,
indicated by the parameter $\delta_{\rm gas}$. As shown by Bruderer
(2013), CO can survive inside the dust cavities in spite of being
exposed to very strong UV radiation thanks to its ability to
self-shield. The modeling demonstrates that a large dynamic range in
gas masses can be probed by CO with ALMA, from 10 Jupiter masses down
to an Earth mass. To quantify the amount of gas, it is best to use a
mix of optically thin and thick CO isotopologs.

As an example of the model results, Figure 2 illustrates the best fit
models for the HD 135344B disk, fitting both the gas and dust images
as well as the spectral energy distribution.  The gas surface density
profile is inferred to drop by 4 orders of magnitude at a cavity
radius $r_{\rm cav, gas}$=30 AU, which is somewhat smaller than the
dust cavity radius at 40 AU.  The dust density drops by an even larger
factor of $>10^4$ at 40 AU.  With the current spatial resolution,
there is some model degeneracy between gas cavity size and value of
$\delta_{\rm gas}$: a larger gas cavity size would imply a shallower
drop.

Similar results are found for the other disks in our sample: in all
cases a drop in the gas density of a factor of 10--100 is found based
on $^{12}$CO data (vdM15a) and a deeper drop is inferred from the CO
isotopologs (vdM15b). The fact that gas is still present inside the
gaps but at a lower level rules out photoevaporation and grain growth
as the main mechanisms for producing the dust cavities and point to
clearing of the cavity by one or more Jovian mass companions.  The
large gas drops are also incompatible with pressure bumps due to
instabilities at dead zone edges as the only cause of the observed
dust traps (Lyra et al.\ 2015).

Modest gas density drops of a factor of 10--100 would suggest that any
embedded planets are unlikely to be more massive than 1--2 Jupiter
masses, and then only for a high disk viscosity (Pinilla et al.\ 2012,
vdM15a). The deeper drops inferred from the CO isotopologs do not
allow to put strong constraints on the characteristics of the embedded
planets unless independent information on the disk viscosity is
available.  The deep drop inferred for IRS 48 together with the vortex
scenario for the azimuthal dust trap (which requires low viscosity)
points to a massive 10--15 M$_{\rm Jup}$ planet in this case.

These data and analyses illustrate the diversity of transitional disk
systems and the potential of ALMA data to infer the characteristics of
embedded planets. While none of the models are unique, the combination
of gas and dust data can rule out certain scenarios. Higher angular
resolution ALMA data at $\sim 0.05''$ resolution of CO isotopologs are
clearly needed to spatially resolve the gaps and quantify the
location, depth and sharpness of the gas drops, which will be a
crucial step to further pin down the models.

\acknowledgements The authors are grateful to K. Dullemond,
T. Birnstiel, T.\ van Kempen, L.\ P\'erez, A.\ Isella, J.\ Brown, K.\
Pontoppidan, V.\ Geers, G.\ Herczeg, G.\ Mathews and S.\ Andrews for
collaborations on these sources. This work is supported by A-ERC grant
291141 CHEMPLAN, the KNAW and NOVA.



\end{document}